# Superior valley polarization and coherence of 2*s* excitons in monolayer WSe$_2$

*Shao-Yu Chen,[1] Thomas Goldstein,[1] Jiayue Tong,[1] Takashi Taniguchi,[2] Kenji Watanabe[2]*

*and Jun Yan[1,\*]*

[1]Department of Physics, University of Massachusetts, Amherst, Massachusetts 01003,

USA

[2]National Institute of Materials Science, 1-1 Namiki, Tsukuba, Ibaraki 305-0044, Japan

[*]Corresponding Author: Jun Yan.　　Tel: (413)545-0853　　Fax: (413)545-1691

E-mail: yan@physics.umass.edu

**Abstract**

We report experimental observation of 2$s$ exciton radiative emission from monolayer tungsten diselenide, enabled by hexagonal boron nitride protected high-quality samples. The 2$s$ luminescence is highly robust and persists up to 150K, offering a new quantum entity for manipulating the valley degree of freedom. Remarkably, the 2$s$ exciton displays superior valley polarization and coherence than 1$s$ under similar experimental conditions. This observation provides evidence that the Coulomb-exchange-interaction-driven valley-depolarization process, the Maialle-Silva-Sham mechanism, plays an important role in valley excitons of monolayer transition metal dichalcogenides.

PACS numbers: 78.55.-m, 78.67.-n, 71.35.Cc

The coupled spin-valley physics [1] in monolayer (1L) transition metal dichalcogenide (TMDC) semiconductors has inspired great strides towards realizing valleytronic devices harnessing these two-dimensional (2D) materials [2–5]. The two energetically degenerate 1L-TMDC valleys with opposite angular momentum can be selectively populated with circularly polarized optical excitation, and the valley polarization can be detected both optically [2–4] and electrically [5]. Further, coherent superposition of valley excitons can be generated with linearly polarized light [6] or a sequence of laser pulses with opposite circular polarization [7], which allows for rotation of the valley pseudospin with magnetic Zeeman effect or optical Stark effect [8,9]. Such coherent manipulations of valley pseudospin are at the heart of future quantum valleytronic devices, and requires thorough understanding and efficient control of various valley depolarization and decoherence processes.

In general, intervalley scattering can occur due to both extrinsic mechanisms such as disorder scattering, and intrinsic mechanisms such as the Coulomb exchange interaction [10]; the competition between these different valley relaxation channels is a topic under active debate [7,11–13]. So far many of the valleytronic studies focus on the $1s$ exciton, the ground state of Coulomb-bound electron-hole pairs, which is readily accessible in 2D TMDC monolayers [2–9,14]. Excitons also have higher energy states that form the hierarchical Rydberg-like series [15–17], similar to a hydrogen atom. It is desirable to access the valley pseudospin of these higher quantum number exciton states, which in previous studies have been employed to demonstrate the exceptionally large exciton binding energy [15–19] and to probe exciton internal quantum transitions [20]. Yet it is relatively challenging to generate radiative emission from these states, as can be understood from Kasha's rule [21]: photon emission quantum yield is appreciable only for the lowest energy excited state, which for the charge neutral exciton, is the $1s$ state. In this Letter, we report that with efficient removal of disorder and phonon scattering channels, the $2s$ exciton luminescence from monolayer tungsten diselenide (1L-WSe$_2$) becomes accessible for valleytronic investigations. This is similar to the breaking of Kasha's rule in high-quality GaAs quantum wells [22], where the $2s$ luminescence becomes observable at low temperatures. We found the 1L-WSe$_2$ $2s$ exciton luminescence to be robust up to 150K, providing a new quantum entity for facile manipulation of valley pseudospins. In contrast to 1*s*, 2*s* exciton exhibits much higher degree of valley polarization and coherence. This observation could be facilitated in part by the fast population decay of 2*s*, and our analysis further points to the action of intervalley Coulomb exchange interaction in TMDC pseudospin propagation, known as the Maialle-Silva-Sham (MSS) mechanism [10], which has been more elusive for charge neutral excitons [11–13] than for trions [6,23,24]. Our studies provide key insights into the TMDC intervalley scattering processes which are essential for developing TMDC-based valleytronic devices.

The 1L-WSe$_2$ samples used in our experiments are mechanically exfoliated from chemical vapor transport grown bulk crystals and are sandwiched between hexagonal boron nitride (*h*BN) flakes using a dry transfer technique (See Supplementary). Figure 1a shows the luminescence and differential reflectance spectra at 20K. In the upper panel, the luminescence spectrum displays a series of sharp peaks with narrow linewidth. The peak at 1.724eV, denoted as $X_{1s}^0$, is the neutral 1*s* exciton. Two peaks around 1.69eV separated by ~7meV are attributed to the coupled intra- and inter-valley trions split by the exchange interaction [6]. In the lower panel, a sharp peak at 1.855eV with full width half maximum (FWHM) of 4.8meV appears and we attribute it to the charge neutral 2*s* exciton luminescence ($X_{2s}^0$). The differential reflectance exhibits two prominent dips that match well to the $X_{1s}^0$ and $X_{2s}^0$ in the luminescence spectra. The near zero luminescence Stokes shift from the absorption dips [25] and the fully resolved negative trion doublet reflect the good sample quality [6,23,24].

Figure 1b shows the temperature dependence of luminescence emission from the sample. Both $X_{1s}^0$ and $X_{2s}^0$ blue shift with narrower linewidths at lower temperatures. In the Supplementary, we have performed detailed fitting and found that the peak position and linewidth evolution of $X_{1s}^0$ and $X_{2s}^0$ can be described by the same formulations. The temperature dependent intensities for the two neutral excitons are plotted in Fig.1c. The $X_{1s}^0$ intensity first increases and then decreases, peaking at about 150K. We note that this is distinct from previous WSe$_2$ samples that display monotonic 1*s* intensity decrease with lowering temperature [26], as a result of disorder scattering that depletes bright excitons into thermal equilibrium with lower energy dark excitons. The excitons in 1L-WSe$_2$ are tightly bound [15] with large wavefunction overlap between the constituent electron and hole, giving rise to large exciton transition dipole oscillator strength and short radiative lifetime [20,27,28]. The non-monotonic 1s intensity temperature dependence is thus a manifestation of out-of-equilibrium exciton radiative recombination becoming more competitive with thermal equilibration between different quantum channels when disorder in the sample is minimized. In contrast, $X_{2s}^0$ does not show up until ~150K and its intensity keeps increasing with lowering temperature. Noting that the 2s-1s exciton energy separation is about 130meV, in the temperature range of our experiment, thermal distribution of the 2s exciton, unlike 1s, is largely negligible. The monotonic increase of 2s intensity at lower temperatures indicates that removal of phonon scattering enhances non-equilibrium 2s radiative emission, and further suggests that the 2s exciton also has fast radiative recombination rate.

We note that there exists some controversy in the assignment of optical features with energies higher than the 1s exciton. Our observed 2s-1s separation of about 130meV is consistent with existing differential reflectance [15] and photoluminescence excitation (PLE) measurements [29], while a separate optical study inferred a much larger 2s-1s separation of 790meV [19]. Optical features in hBN sandwiched WSe$_2$ heterostructures are further complicated by inter-material exciton-phonon coupling that results in hybrid modes that do not appear in the optical spectra of either hBN or WSe$_2$ alone [30,31]. To confirm that the new emission feature we observe is from the 2s exciton, we performed two more control experiments. First, we fabricated an hBN-sandwiched field effect transistor device to tune this new peak by charge doping. We found that both $X_{1s}^0$ and $X_{2s}^0$ radiation become weak and eventually disappear when the crystal is doped with electrons or holes (Supplementary Fig.S3). This confirms that both $X_{1s}^0$ and $X_{2s}^0$ are associated with neutral excitons, consistent with our assignment. Second, we tuned the laser excitation across the $X_{2s}^0$ energy range to perform one photon PLE and resonant Raman scattering measurements. The $X_{1s}^0$ luminescence becomes more intense when the incident photon is in resonance with the $X_{2s}^0$ energy (Supplementary Fig.S4). Further, two Raman bands R$_1$ and R$_2$ at 128 and 132meV become visible in Fig.2a, consistent with another recent Raman study that found a broad phonon feature in the range of 128-133meV (1030-1070cm$^{-1}$) [30]. These two bands are assigned as the combinational modes [30,31] arising from the out-of-plane vibrations of WSe$_2$ (OC: <u>o</u>ut-of-plane <u>c</u>halcogen vibration [32], 31meV) and hBN (ZO: <u>z</u>-direction <u>o</u>ptical phonon; the infrared active 97meV A$_{2u}$ [33] and the optically silent 101meV $B_{1g}$ [34] phonons). The $R_1$ and $R_2$ bands have energies that are quite close to the 1s-2s energy separation; one possibility is that the $X_{2s}^0$ emission we observe are $R_1$ and $R_2$ phonon-exciton replicas of $X_{1s}^0$. We rule out that interpretation through two observations. One, as can be seen in Fig.2a, the combinational phonon bands are composed of two distinct peaks separated by ~4meV with a non-symmetric lineshape that depends sensitively on the resonance condition, while the $X_{2s}^0$ emission spectrum can be well-fitted by a Lorentzian function (Fig.1). Two, we measured the temperature dependence of the combinational phonon bands (Fig.2b) and found that the energy shift is opposite to that of the 1s-2s separation (Fig.2c and Supplementary Fig.S5). This confirms that the $X_{2s}^0$ emission is not related to $R_1$ and $R_2$.

The appearance of the $X_{2s}^0$ emission in high-quality samples allows us to examine its valleytronic properties. Taking advantage of the valley dependent optical selection rule [1], we use circularly polarized light to selectively populate one valley and monitor the resultant valley polarization by examining the helicity of optical emission [2–4]. We also use linearly polarized light to create a coherent superposition of excitons in both K and K' valleys; the decoherence of the valley excitons are reflected in the degree of linear polarization of the luminescence emission [6]. Experimentally we excite our sample at 20K with $\sigma_+$ circularly polarized and $H$ linearly polarized laser light that is detuned by 20meV above the exciton energy, and collect the luminescence emission with $\sigma_+$, $\sigma_-$, $H$ and $V$ polarizations; see Fig.3a. The valley polarization and coherence are characterized by $P = \frac{I_{\sigma+\sigma+}-I_{\sigma+\sigma-}}{I_{\sigma+\sigma+}+I_{\sigma+\sigma-}}$ and $C = \frac{I_{HH}-I_{HV}}{I_{HH}+I_{HV}}$ respectively.

From Fig.3a, we found the 2s excitons to exhibit superior capability in retaining the broken time reversal symmetry and coherence of incident laser light with $P = 0.82$ and $C = 0.56$. Similar measurements are performed for the 1s exciton; see right panel of Fig.3a. Interestingly its $P = 0.15$ and $C = 0.17$ are significantly smaller than 2s, although the measurement was performed in the same sample at the same temperature with the laser energy also detuned at 20meV above the exciton.

The superior 2s valley polarization could be assisted by its fast population decay rate. As a higher energy state, the 2s exciton possesses decay channels such as the 2s-1s transition (See Supplementary) not available to 1s. Indeed, $X_{2s}^0$ has a wider linewidth than $X_{1s}^0$ (4.8 vs. 4.0meV at 20K, see Fig.1a). If we assume the 0.8meV linewidth difference is mostly due to faster population decay, and take the 1s luminescence decay time to be 2ps from a recent study [35], we infer a 2s lifetime of about 0.6ps. Further the 2s oscillator strength is about 15 times weaker than 1s from absorption spectra in Fig.1, consistent with the value from a recent diamagnetic shift measurement [36], while the low temperature 1s intensity is about 60 times stronger than 2s (Fig.1c). This suggests a decay rate ratio of 4, in reasonable agreement with the above estimation from linewidth difference. Assuming a phenomenological relation between P, the population and polarization decay time $\tau$ and $\tau_s$: $P = \frac{1}{1+\tau/\tau_s}$, and using P = 0.15 and 0.82 for 1s and 2s, we find $\tau_s$ is about six times larger for 2s than for 1s. This indicates that the 2s exciton valley polarization is intrinsically more robust than 1s. Noting that the 2s and 1s excitons have the same symmetry, intervalley scattering allowed for 1s is thus anticipated to also affect the 2s valley pseudospins. Quantitatively however, the scattering rates may differ. In particular, the exchange interaction, capable of inducing intrinsic valley depolarization and decoherence through the MSS mechanism [10], differs substantially for 1s and 2s excitons. A recent study showed that MSS plays an important role in valley decoherence and observed a coherence time of about 100fs [7]. Below, we explain the drastically different valley polarization and coherence for 1s and 2s excitons in the framework of the exchange interaction MSS mechanism.

As illustrated in Fig.3b, the strong Coulomb interaction between the photo-generated electrons and holes not only gives rise to exceptionally large exciton binding energy [15], but also leads to the annihilation of bright excitons in one valley and creation in the other. This exchange of the excitons between the two valleys conserves energy but induces flipping of exciton angular momentum and pseudospin, compromising the valley polarization and coherence. For excitons with center-of-mass momentum $\vec{k}$, the intervalley exchange interaction is given by [37]

$$J_{\vec{k}} = -|\psi(r_{eh}=0)|^2 \frac{a^2 t^2}{E_g^2} V(\vec{k}) k^2 e^{-2i\theta} \qquad (1)$$

where $\psi(r_{eh})$ is the real space wavefunction for the relative motion between the electron and the hole, $a = 3.32$Å is the lattice constant of monolayer WSe$_2$, $t$=1.19eV is the hopping energy, $E_g \approx$ 2eV is the band gap, $V(\vec{k})$ is the $\vec{k}$ component of the Coulomb interaction, and $\theta$ denotes the direction of $\vec{k}$. Effectively this exchange interaction introduces a pseudo-magnetic field acting on the valley pseudospin of the excitons. The angular dependence in Eqn.(1) implies that the direction of the pseudo-magnetic field depends on the direction of the exciton wavevector (Fig.3c). Consider, for example, a set of excitons with the same energy and pseudospin populated on a ring in the $\vec{k}$ space. The pseudo magnetic fields acting on them will have the same magnitude but different directions depending on the direction of $\vec{k}$. This makes the excitons on the ring to precess towards different directions, which in turn, causes valley depolarization and decoherence as the excitons propagate.

In Eqn.(1), $|\psi(r_{eh} = 0)|^2$ describes the probability density for the electron and the hole to spatially overlap. For the 1s exciton this is given approximately by $1/a_B^2$, where $a_B$ ≈1.7nm [36] is the exciton Bohr radius. In the case of 2s excitons, a recent measurement found that the electron-hole separation in 2s is about 6.6nm [36]. Assuming that the 1s and 2s excitons have about the same mass, the 2s exchange interaction is then about 15 times weaker. This difference has an important impact on the exciton valley pseudospin dynamics. In Fig.3d, we simulated the pure exchange-interaction-driven valley depolarization and decoherence for excitons with different momentum $k$ and kinetic energy $E_k = k^2/2M$: at $k$ = 0, both $P$ and $C$ are equal to 1 since the exchange interaction in Eqn.(1) goes to zero at $k$ = 0; for nonzero $k$, both $P$ and $C$ of 1s drops steeply at finite $E_k$, while for 2s the decrease is much slower, confirming that 1s is more impacted by the exchange depolarization fields.

It is of interest to note that for both 1s and 2s simulations in Fig.3d, $C$ is always larger than $P$ — this is a hallmark of exciton exchange interaction in 2D [10]: the exchange-interaction-induced pseudo-magnetic-fields are in the plane of the atomic layer, thus the out-of-plane pseudospin of valley polarized excitons experiences the pseudo magnetic fields in two directions, while the in-plane pseudospin of the valley coherent excitons is relaxed only by the magnetic field component that is perpendicular to the pseudospin. Experimentally, we have observed $C$ to be larger than $P$ for 1s in Fig.3a as well as with many other laser excitations (more data in Supplementary Fig.S4), further confirming that the exchange interaction dominates the 1s exciton valleytronic behavior. This is consistent with another recent study on high-quality $MoS_2$ where $C$ is also found to be larger than $P$ [38].

We note that for 2s excitons however, *P* is significantly larger than *C* as shown in Fig.3a. This suggests that with weaker 2s exchange interaction, other decoherence and depolarization mechanisms become more competitive. To account for these additional mechanisms, we have modified the model (see Supplementary) such that even for $k = 0$, *P* and *C* are smaller than 1. This relatively simple model captures our observations semi-quantitatively: as shown in Fig.3e, for excitons with small kinetic energy ($E_k <$ 1meV), *P* is mostly larger than *C* for 2s and smaller than *C* for 1s, and numerically the 2s *P* and *C* values are much larger than 1s.

We finally remark that the excitons can only become radiative if its momentum lies within the light cone, whose boundary corresponds to 1s and 2s exciton kinetic energy of ~10μeV. At such small $E_k$'s the impact of exchange interaction is small. The large difference between *P* and *C* for 1s and 2s agrees with the conjecture that excitons outside the light cone with larger momentum provide a reservoir where disorder and phonon can scatter them into the light cone, which subsequently radiate [20]. The average exchange interaction that the radiatively recombined excitons experienced is thus much larger than the fields inside the light cone. In the supplementary, we show that it is possible to reduce the impact of exchange interaction fields on 1s by using the small-momentum 2s exciton as an alternative reservoir, corroborating another study of WSe$_2$ on SiO$_2$ [39]. Here with the presence of *h*BN, the 2s exciton can lose the excess ~130meV by emitting zero-momentum *h*BN-WSe$_2$ combinational phonons (Fig.2a). This reduces the number of phonons involved from six [39] to two, and markedly improves the 1s valley coherence and polarization to 0.64 and 0.30 respectively (Supplementary Fig.S4).

In conclusion, we have accessed the 2s radiative emission in *h*BN sandwiched high-quality 1L-WSe$_2$ crystals. The 2s luminescence is highly robust and exhibits superior valleytronic properties. Our data provide evidence that the Maialle-Silva-Sham mechanism plays an importance role in the exciton valley decoherence and depolarization, which should be taken into account when developing valleytronic devices.

**Acknowledgements**

We thank Tony Heinz, Jie Shan and Kin Fai Mak for helpful discussions. This work is supported mainly by the University of Massachusetts Amherst, and in part by NIST 60NANB12D253 and NSF ECCS-1509599. T.T. and K.W. acknowledge support from the Elemental Strategy Initiative conducted by the MEXT, Japan and JSPS KAKENHI Grant Numbers JP26248061, JP15K21722 and JP25106006.

**Note:** Recently we became aware of a related work that also observed the 2$s$ exciton luminescence in high-quality WSe$_2$ monolayers [40].

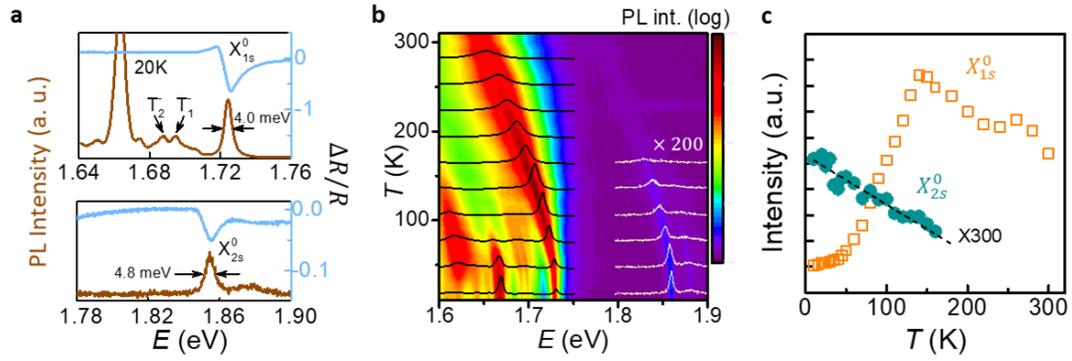

**Figure 1** (**a**) The photoluminescence (brown) and differential reflectance (light blue) spectra at 20K. The FWHM of 1$s$ ($X_{1s}^0$) and 2$s$ ($X_{2s}^0$) are 4.0 and 4.8meV, respectively. (**b**) Photoluminescence spectra plotted as a function of temperature. Selected spectra at $T$ = 10 to 280K with 30K steps are displayed. (c) Temperature dependences of 1$s$ and 2$s$ intensity.

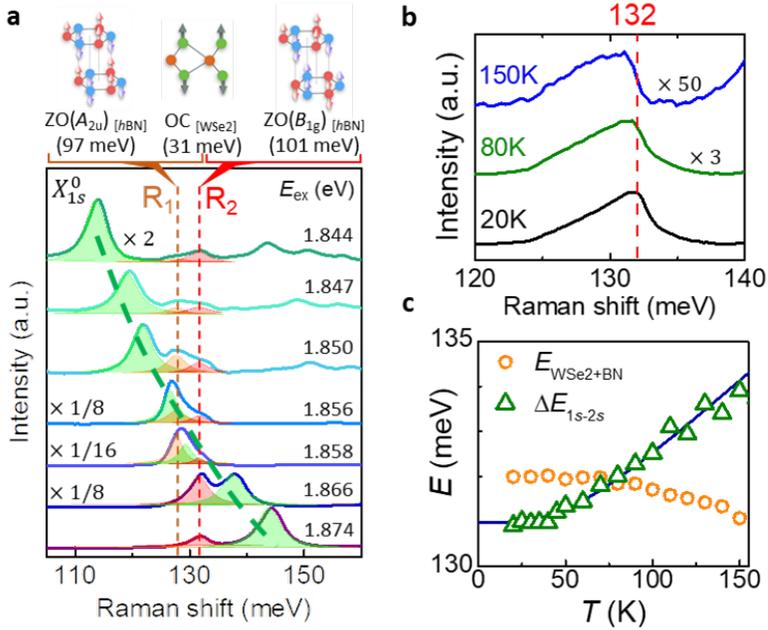

**Figure 2** (**a**) Resonant Raman scattering of $R_1$ and $R_2$ using photon energies from 1.844 to 1.874eV. The peaks guided by dashed curve are the 1s exciton luminescence. (**b**) Raman scattering of the WSe$_2$/BN combinational modes at 20K, 80K and 150K. The dash line is aligned with 132meV. (**c**) The temperature dependent energy of 1s and 2s excitons separation ($\Delta E_{1s-2s}$) and WSe$_2$/BN combinational phonons.

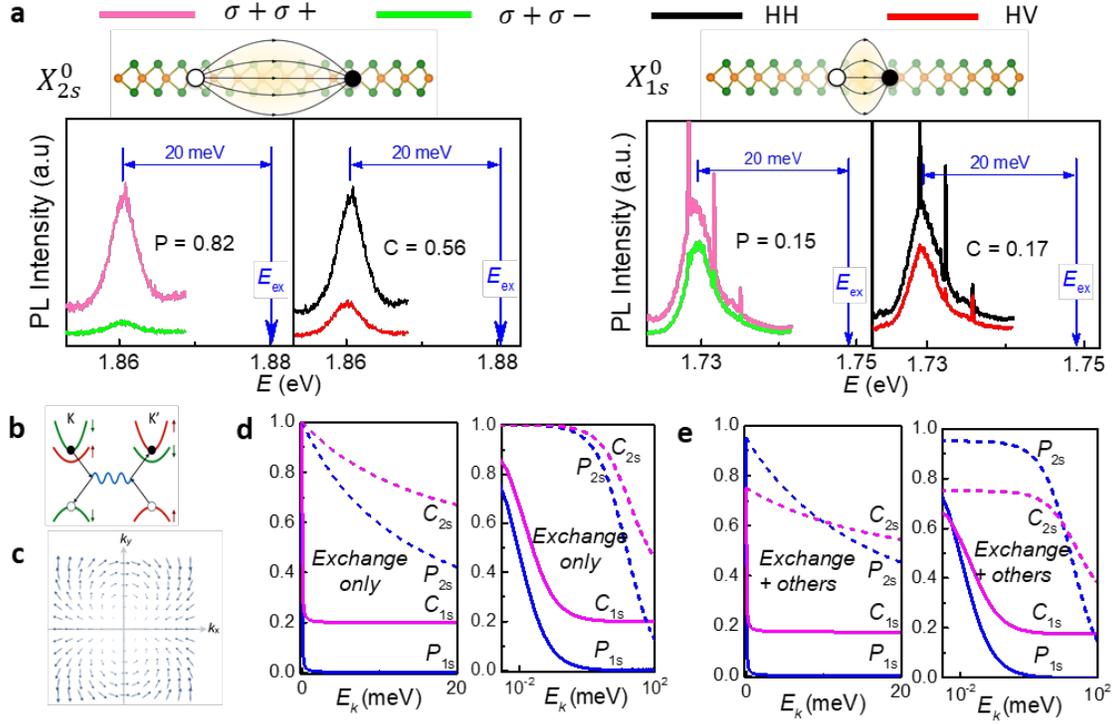

**Figure 3 (a)** The circular and linear polarization-resolved photoluminescence of 1L-WSe$_2$ at 20K with detuned excitation photon energy at 20meV above 2$s$ (left) and 1$s$ (right) excitons. **(b)** A schematic showing the inter-valley electron-hole exchange interaction, which induces pseudospin flip. **(c)** The strength and direction of the inter-valley exchange pseudo-magnetic field in $k$-space. **(d)** The simulated valley coherence ($C$) and polarization ($P$) as a function of $E_k$ for 1$s$ and 2$s$ excitons considering pure exchange interactions. The left (right) panel is in linear (semilog) scale. **(e)** Simulated $C$ and $P$ considering both exchange interactions and other depolarization and decoherence mechanisms.

## 1. Sample fabrication and optical measurement setup

The high-quality monolayer tungsten diselenide (1L-WSe$_2$) samples are exfoliated from bulk WSe$_2$ crystals grown using the chemical vapor transport method. High purity W 99.99%, Se 99.999%, and I$_2$ 99.99% (Sigma Aldrich) are placed in a fused silica tubing that is 300 mm long with an internal diameter of 18 mm. The temperatures are set at 1055°C for the reaction zone and at 955°C for the growth zones. Optical microscopy and atomic force microscopy are used to select clean and flat 1L-WSe$_2$, few layer $h$BN and few layer graphene. The atomic flakes exfoliated on 300 nm SiO$_2$/Si wafers are stacked using a dry transfer technique with PPC(poly-propylene carbonate) stamps [1]. Figure S1 displays the image of a BN/1L-WSe$_2$/BN heterostructure taken by differential interference contrast enhanced optical microscope.

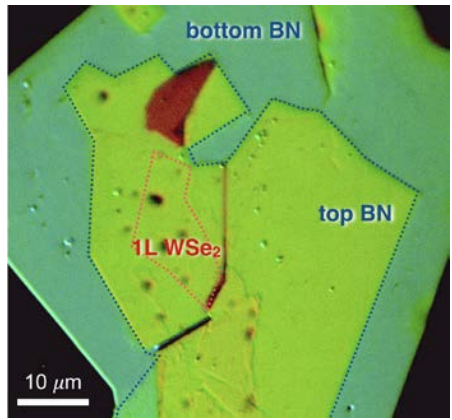

**Figure S1.** Optical Microscope image of a BN/1L-WSe$_2$/BN heterostructure.

After fabrication, the sample is transferred to a microscopy cryostat with a base pressure of $1.0 \times 10^{-6}$ Torr for spectroscopy measurements. The linear polarization and helicity resolved optical setup is similar to the ones used in our previous works [2–5]. In this paper, we employ three types of laser excitations: a frequency doubled Nd:YAG solid state laser (532nm), a dye laser (612-672nm) and a Ti:Sapphire laser (709nm-920nm). The differential reflectance is performed with a supercontinuum white laser from NKT photonics [5]. The incident light is focused on the sample by a 40× objective lens (NA: 0.6) with a spot size of ~2 $\mu$m. The power of excitation laser for all measurements is kept below 100 $\mu$W to minimize heating effects. The light signal is detected by a triple spectrometer (Horiba T64000) equipped with a liquid nitrogen cooled CCD camera.

2. **Temperature and electrostatic doping dependence of 1s and 2s excitons**

The temperature dependence of intensity is shown in main text Fig.1c. Here we carry out analysis on the 1s and 2s peak energy and linewidth. As can be seen in Fig.S2a and S2b, the $X_{1s}^0$ and $X_{2s}^0$ peak energies and linewidths show similar temperature dependence, which can be fitted to the same models of hyperbolic cotangent relation (Eq. S1) [6] and phonon induced broadening (Eq. S2) [7] respectively:

$$E_g(T) = E_0 - S\langle\hbar\omega\rangle\left[\coth\left(\frac{\langle\hbar\omega\rangle}{2k_BT}\right) - 1\right] \quad (S1)$$

where $E_0$ is the optical bandgap at $T = 0K$, $S$ is the coupling factor, and $\langle\hbar\omega\rangle$ represents the average phonon energy in the system [8];

$$\gamma = \gamma_0 + c_1 T + \frac{c_2}{e^{\frac{\langle\hbar\omega\rangle}{k_BT}} - 1} \quad (S2)$$

where $\gamma_0$ is the FWHM at $T = 0K$, the second and third terms account for the impacts of acoustic and optical phonons respectively, and $\langle\hbar\omega\rangle$ matches well with the zone-center out-of-plane chalcogen (OC) and in-plane metal-chalcogen (IMC) vibrations which are accidentally degenerate in 1L-WSe$_2$ [2]. From the fitting parameters (Table S1), $\gamma_0$ is $3.80 \pm 0.05$meV, in agreement with the intrinsic FWHM of $3.8 \pm 0.4$meV measured with four wave mixing at 5K [9].

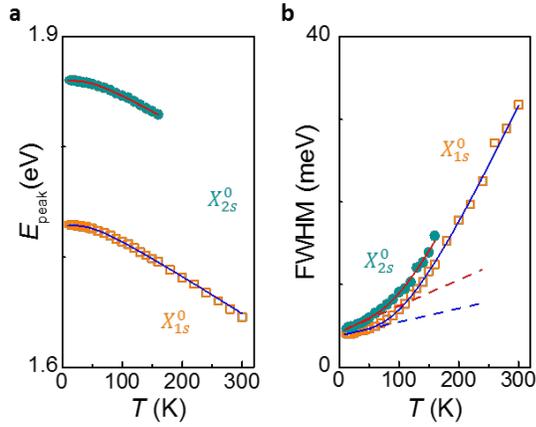

**Figure S2**. Temperature dependent peak energy **(a)** and linewidth **(b)** of $X_{1s}^0$ and $X_{2s}^0$ radiations. The solid curves are fits using equations S1 and S2. The dashed lines in **(b)** represent the linear terms which dominate at low temperatures.

**Table S1.** The fitting parameters for the temperature dependent peak energy and linewidth of 1s and 2s exciton luminescence.

|  | peak energy | | | linewidth | | | |
|---|---|---|---|---|---|---|---|
|  | $\langle \hbar\omega \rangle$ | $E_0$ | $S$ | $\langle \hbar\omega \rangle$ | $\gamma_0$(FHWM) | $c_1$ | $c_2$ |
|  | $meV$ | $eV$ |  | $meV$ | $meV$ | $\mu eV/K$ |  |
| $X^0_{1s}$ | 13.0 ± 0.4 | 1.728 | 2.01 ± 0.03 | 31 ± 1 | 3.80 ± 0.05 | 16.0 ± 1.4 | 0.05 ± 0.002 |
| $X^0_{2s}$ |  | 1.860 | 1.85 ± 0.05 |  | 4.20 ± 0.06 | 34.0 ± 1.7 | 0.05 ± 0.004 |

We also performed electrostatic doping dependent PL measurements using an $h$BN-sandwiched field effect transistor device to understand the relation between $X^0_{1s}$ and $X^0_{2s}$ emissions. Figure S3 shows the PL spectra as a function of gate voltage at 78K. We found that the intensity of $X^0_{1s}$ and $X^0_{2s}$ are highly correlated and only appear at low doping, indicating that both $X^0_{1s}$ and $X^0_{2s}$ are associated with neutral excitons.

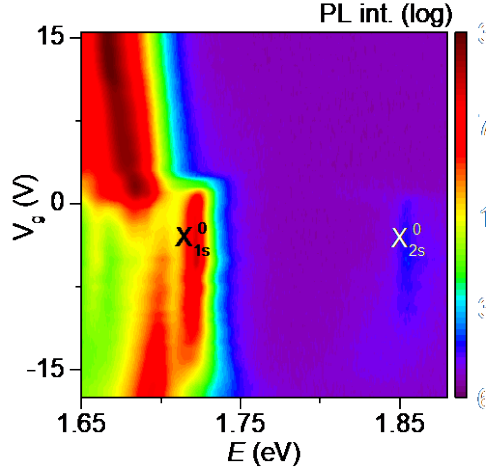

**Figure S3**. PL spectra plotted as a function of gate voltage for a WSe$_2$ field effect transistor device.

### 3. Photoluminescence excitation (PLE) of 1s exciton

In the main text Fig.3 we showed $C$ and $P$ for 1s and 2s with laser excitation at 20 meV above the exciton energy. Here we show additional data for 1s with laser excitation from 1.84 to 1.89eV, covering the energy range of the 2s exciton (Fig.S4a) to explore the impact of 2s-1s transition. In Fig.S4b the 1s luminescence emission is plotted such that its peak position corresponds to the laser excitation energy. Figure S4c displays the experimental $C$ and $P$ values. We observe that $C$ is larger than $P$ over the whole PLE range, confirming that exchange interaction dominates the 1$s$ valley depolarization and decoherence as discussed in the main text.

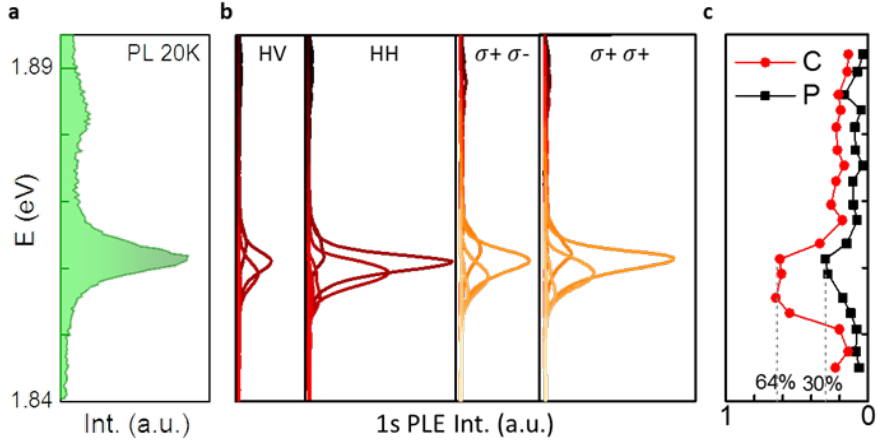

**Figure S4** (**a**) The photoluminescence spectrum of 2$s$ exciton at 20K excited by a 2.33 eV laser. (**b**) 1$s$ exciton luminescent emission plotted such that the peak position corresponds to the energy of the laser excitation. (**c**) $C$ and $P$ of the 1$s$ exciton as a function of laser excitation energy.

In Fig.S4 when the excitation is off resonance from the 2$s$ exciton, $P$ and $C$ are in the range of 0.1 to 0.2. When on-resonance the $P$ and $C$ are markedly improved to 0.30 and 0.64 respectively. This improvement results from using the small-momentum 2$s$ exciton as a high-quality reservoir for the 1$s$ luminescence. The 2$s$ exciton reservoir is prepared by illuminating the sample with photons that match the 2$s$ exciton energy. The reservoir excitons subsequently lose ~130meV excess energy through phonon emission. Our PLE study is similar to a previous work on monolayer $WSe_2$ deposited on Si/SiO$_2$ substrate [10]. In Ref. [10], it was proposed that 6 $WSe_2$ phonons are emitted during the 2$s$-1$s$ relaxation. Here thanks to the $h$BN-$WSe_2$ interlayer exciton-phonon interactions, we can take advantage of not only $WSe_2$ but also $h$BN zone-center phonons. This enabled us to reduce the number of zone-center phonons involved into two, as shown in Fig.2a of the main text. This may explain why our maximum $C$ of 0.64 is much larger than those in Ref. [10].

We note that quantitatively, while our observed maximum $C$ of 0.64 is among the highest in literature [10–13], our maximum $P$ of 0.3 for 1s is smaller than some other samples with more disorder broadening [11,14]. This suggests that $C$ is quite sensitive to disorder dephasing, while $P$ might be improved by motional narrowing [15]. More systematic studies are needed to elucidate the impact of disorder on $P$ and $C$ of 1L-TMDCs.

4. **Temperature dependence photoluminescence, differential reflection and WSe$_2$/BN combinational phonon mode.**

In this section, we provide detailed comparison between the 2s exciton and the R$_1$ and R$_2$ phonon bands. In Fig.S5, we plot 2s exciton PL, differential reflectance and the WSe$_2$/BN combinational phonons at $T$ = 20K, 50K, 80K, 110K and 150K; see panels a,b and c, respectively. For easy comparison, the 1s exciton energy has been subtracted from the PL and the differential reflectance spectra. The dash arrows indicate the directions of the energy shifts with temperature change. As can be seen, the PL and the differential reflectance exhibit similar trends: the energy increases as temperature goes up. In contrast, the WSe$_2$/BN combinational phonons show the opposite trend. This observation further indicates that both the emission peak and the absorption dip in lower panel of Fig.1a in the main text are associated with $X_{2s}^0$, rather than the Raman bands R$_1$ and R$_2$, substantiating our conclusion that the emission we observe at 1.855eV at 20K is not due to phonon replicas of the 1s exciton.

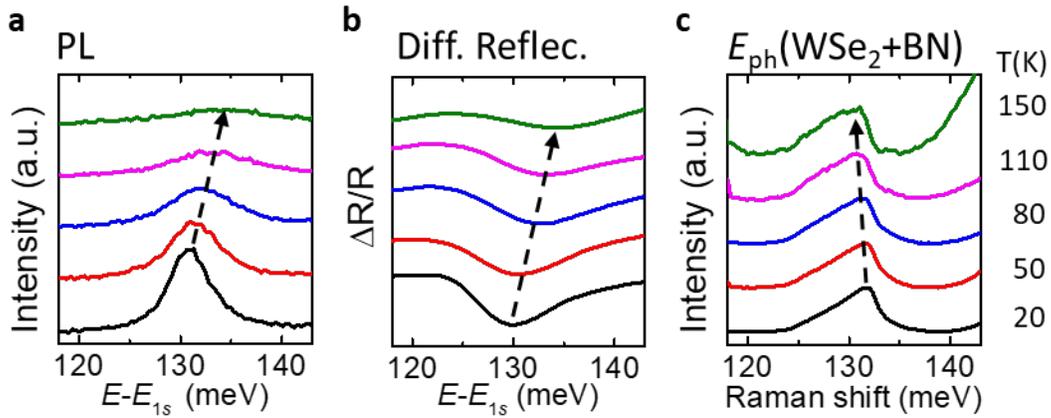

**Figure S5** Representative spectra of **(a)** 2s exciton photoluminescence **(b)** 2s exciton differential reflectance and **(c)** the WSe$_2$/BN combinational phonon at various temperatures. For **(a)** and **(b)** the x-axis is the energy difference from 1s exciton; for **(c)** the x-axis is the energy shift from the laser excitation energy. The dash arrows indicate the directions of energy shift with temperature change.

5. **Valley depolarization and decoherence model.**

The large exciton binding energy in WSe$_2$ from direct Coulomb attraction implies large impacts of exchange interactions. For the bright 1s and 2s exciton branches under consideration, the electron-hole exchange interaction can occur both within the same valley and in-between the two inequivalent valleys. The intravalley exchange gives an overall shift of the exciton energy in both valleys while the intervalley exchange leads to valley mixing. The Hamiltonian that governs the exciton equation of motion can be written as a 2×2 matrix operator acting on the valley pseudo-spin Hilbert space [16]:

$$H(k) = \begin{bmatrix} E_0 + \dfrac{\hbar^2 k^2}{2M} & J(k)e^{-i2\theta} \\ J(k)e^{i2\theta} & E_0 + \dfrac{\hbar^2 k^2}{2M} \end{bmatrix}$$

where $E_0$ is zero-momentum exciton energy determined by the bare bandgap, direct Coulomb interaction, and intravalley exchange interaction, $k$ is the center of mass wavenumber, $\dfrac{\hbar^2 k^2}{2M}$ is the exciton center of mass kinetic energy, $J(k) = -|\psi(r_{eh} = 0)|^2 \dfrac{a^2 t^2}{E_g^2} V(k) k^2$ is the exchange interaction, $V(k) = \dfrac{2\pi e^2}{4\pi\epsilon_0 \epsilon k}$ is the $k$ component of the Coulomb interaction, and $\theta$ characterizes the direction of the exciton center of mass momentum. The exciton mass is $M \approx 0.8 m_e$; we take the effective electron-hole separation as 1.7nm and 6.6nm for 1s and 2s respectively [17]; $a = 3.32$Å is the lattice constant of monolayer WSe$_2$; $t = 1.19$ eV is the hopping energy; $E_g \approx 2$ eV is the band gap; $\epsilon \approx 5$ is the dielectric constant of $h$BN [18–23].

The exciton dynamics is governed by the above Hamiltonian, plus terms relating to scattering between different momentum states, decay due to recombination, and external generation [15]. Written in terms of the density matrix in the pseudospin basis:

$$\frac{d}{dt}\bar{\rho}(\vec{k},t) = \frac{i}{\hbar}[\bar{\rho}(\vec{k},t), H(\vec{k})] + \sum_{\vec{k}'} W_{\vec{k}\vec{k}'}[\bar{\rho}(\vec{k}',t) - \bar{\rho}(\vec{k},t)] - \frac{\bar{\rho}(\vec{k},t)}{\tau} + \bar{G}(\vec{k},t).$$

Writing the exciton density matrix as $\bar{\rho}(\vec{k},t) = \frac{1}{2}N(\vec{k},t)\sigma_0 + \frac{1}{2}\vec{S}(\vec{k},t)\cdot\vec{\sigma}$ and the generation matrix as $\bar{G}(\vec{k},t) = G_N(\vec{k},t)\frac{\sigma_0}{2} + \vec{G}(\vec{k},t)\cdot\frac{\vec{\sigma}}{2}$, with $\vec{\sigma}$ the tensor of Pauli matrices, the evolution of exciton population $N(\vec{k},t)$ and vector $\vec{S}(\vec{k},t)$ are given by:

$$\frac{d}{dt}N(\vec{k},t) = \sum_{\vec{k}'} W_{\vec{k}\vec{k}'}[N(\vec{k}',t) - N(\vec{k},t)] - \frac{N(\vec{k},t)}{\tau} + G_N(\vec{k},t),$$

$$\frac{d}{dt}\vec{S}(\vec{k},t) = \vec{\Omega}(\vec{k}) \times \vec{S}(\vec{k},t) + \sum_{\vec{k}'} W_{\vec{k}\vec{k}'}[\vec{S}(\vec{k}',t) - \vec{S}(\vec{k},t)] - \frac{\vec{S}(\vec{k},t)}{\tau} + \vec{G}(\vec{k},t),$$

where $\vec{\Omega}(\vec{k}) = \frac{i}{\hbar}J(k)[\cos(2\phi),\sin(2\phi),0]$.

Performing Fourier transformations $N(\vec{k},t) = \sum_n N^n(k,t)e^{in\phi}$, $\vec{S}(\vec{k},t) = \sum_n \vec{S}^n(k,t)e^{in\phi}$, and accounting here only elastic scattering between different momentum states (ad-hoc impacts of inelastic scattering on depolarization and decoherence are added later), i.e.: $W_{\vec{k}\vec{k}'} = W(\phi - \phi') = \sum_n W_n e^{in(\phi-\phi')}$, we find that the time dependence of $S$ is block diagonal in the basis of Fourier components. Letting $S_\pm^n = S_x^n \pm iS_y^n$ and $\hbar(W_0 - W_n) = \frac{\hbar}{\tau_n}$, we find:

$$\frac{dN^0(k,t)}{dt} = -\frac{N^0(k,t)}{\tau} + G_N^0(k,t)$$

$$\hbar\frac{d}{dt}\begin{bmatrix}S_z^0\\S_-^{-2}\\S_+^2\end{bmatrix} = \begin{bmatrix}-\frac{\hbar}{\tau} & iJ(k) & -iJ(k)\\2iJ(k) & -\frac{\hbar}{\tau}-\frac{\hbar}{\tau_2} & 0\\-2iJ(k) & 0 & -\frac{\hbar}{\tau}-\frac{\hbar}{\tau_2}\end{bmatrix}\begin{bmatrix}S_z^0\\S_-^{-2}\\S_+^2\end{bmatrix} + \hbar\begin{bmatrix}G_z^0\\G_-^{-2}\\G_+^2\end{bmatrix}$$

$$\hbar \frac{d}{dt}\begin{bmatrix} S_+^0 \\ S_z^{-2} \\ S_-^{-4} \end{bmatrix} = \begin{bmatrix} -\frac{\hbar}{\tau} & -2iJ(k) & 0 \\ -iJ(k) & -\frac{\hbar}{\tau}-\frac{\hbar}{\tau_2} & iJ(k) \\ 0 & 2iJ(k) & -\frac{\hbar}{\tau}-\frac{\hbar}{\tau_4} \end{bmatrix} \begin{bmatrix} S_+^0 \\ S_z^{-2} \\ S_-^{-4} \end{bmatrix} + \hbar \begin{bmatrix} G_+^0 \\ G_z^{-2} \\ G_-^{-4} \end{bmatrix}$$

where $S_z^0$ and the real part of $S_+^0$ describe the evolution of valley polarization and coherence respectively. The solutions to these dynamic equations are given by exponentially decaying terms and steady terms of the form $G$ divided by an appropriate time constant. In our continuous wave measurements only the steady terms are relevant, and the final distribution of $S$ and $N$ can be solved analytically.

For calculation of depolarization, we generate excitons with $\bar{G}_P(k) = G_P(k)\begin{bmatrix} 1 & 0 \\ 0 & 0 \end{bmatrix} = \frac{G_P(k)}{2}\sigma_0 + \frac{G_P(k)}{2}\sigma_z$. The steady state polarization is given by

$$P = \frac{I_{\sigma+\sigma+} - I_{\sigma+\sigma-}}{I_{\sigma+\sigma+} + I_{\sigma+\sigma-}} = \frac{Tr\left(\begin{bmatrix}1 & 0\\0 & 0\end{bmatrix}\cdot\bar{\rho}\right) - Tr\left(\begin{bmatrix}0 & 0\\0 & 1\end{bmatrix}\cdot\bar{\rho}\right)}{Tr\left(\begin{bmatrix}1 & 0\\0 & 0\end{bmatrix}\cdot\bar{\rho}\right) + Tr\left(\begin{bmatrix}0 & 0\\0 & 1\end{bmatrix}\cdot\bar{\rho}\right)} = \frac{S_z}{N} = \frac{1}{1+\frac{4J(k)^2}{\frac{\hbar}{\tau}\left(\frac{\hbar}{\tau}+\frac{\hbar}{\tau_2}\right)}}.$$

For calculation of valley coherence, we generate excitons with $\bar{G}_C(k) = G_C(k)\begin{bmatrix} \frac{1}{2} & \frac{1}{2} \\ \frac{1}{2} & \frac{1}{2} \end{bmatrix} = \frac{G_C(k)}{2}\sigma_0 + \frac{G_C(k)}{2}\sigma_x$. The steady state coherence is given by

$$C = \frac{I_{HH} - I_{HV}}{I_{HH} + I_{HV}} = \frac{Tr\left(\begin{bmatrix}\frac{1}{2} & \frac{1}{2}\\ \frac{1}{2} & \frac{1}{2}\end{bmatrix}\cdot\bar{\rho}\right) - Tr\left(\begin{bmatrix}\frac{1}{2} & -\frac{1}{2}\\ -\frac{1}{2} & \frac{1}{2}\end{bmatrix}\cdot\bar{\rho}\right)}{Tr\left(\begin{bmatrix}\frac{1}{2} & \frac{1}{2}\\ \frac{1}{2} & \frac{1}{2}\end{bmatrix}\cdot\bar{\rho}\right) + Tr\left(\begin{bmatrix}\frac{1}{2} & -\frac{1}{2}\\ -\frac{1}{2} & \frac{1}{2}\end{bmatrix}\cdot\bar{\rho}\right)} = \frac{Re(S_+)}{N} = \frac{2J(k)^2 + \left(\frac{\hbar}{\tau}+\frac{\hbar}{\tau_2}\right)\left(\frac{\hbar}{\tau}+\frac{\hbar}{\tau_4}\right)}{2J(k)^2\left(2+\frac{\tau}{\tau_4}\right) + \left(\frac{\hbar}{\tau}+\frac{\hbar}{\tau_2}\right)\left(\frac{\hbar}{\tau}+\frac{\hbar}{\tau_4}\right)}.$$

In the simulation we used $\frac{\hbar}{\tau} = 0.33$meV which corresponds to 1s exciton population decay rate of 2ps [24]. For $\frac{\hbar}{\tau_2}$ and $\frac{\hbar}{\tau_4}$ 1meV is used. In the case of 2s, as discussed in the main text, we estimated its decay rate to be 3-4 times faster than 1s, and a population decay time of 0.5ps is used.

To take into account depolarization and decoherence mechanisms other than the exchange interaction, we modify the spin dynamic equation such that

$$\frac{d}{dt}\vec{S}(\vec{k},t) = \vec{\Omega}(\vec{k}) \times \vec{S}(\vec{k},t) + \sum_{\vec{k}'} W_{\vec{k}\vec{k}'}[\vec{S}(\vec{k}',t) - \vec{S}(\vec{k},t)] - \begin{bmatrix} \frac{1}{\tau_c} & 0 & 0 \\ 0 & \frac{1}{\tau_c} & 0 \\ 0 & 0 & \frac{1}{\tau_p} \end{bmatrix} \vec{S}(\vec{k},t) + \vec{G}(\vec{k},t).$$

This gives $P = \dfrac{\tau_p/\tau}{1+\frac{4J(k)^2}{\frac{\hbar}{\tau_p}\left(\frac{\hbar}{\tau_c}+\frac{\hbar}{\tau_2}\right)}}$ and $C = \dfrac{\tau_c}{\tau}\dfrac{2J(k)^2+\left(\frac{\hbar}{\tau_p}+\frac{\hbar}{\tau_2}\right)\left(\frac{\hbar}{\tau_c}+\frac{\hbar}{\tau_4}\right)}{2J(k)^2\left(2+\frac{\tau_c}{\tau_4}\right)+\left(\frac{\hbar}{\tau_p}+\frac{\hbar}{\tau_2}\right)\left(\frac{\hbar}{\tau_c}+\frac{\hbar}{\tau_4}\right)}$. Note that $\tau_c < \tau_p < \tau$ and for our simulation in Fig.3e we have set $\frac{\tau_p}{\tau} = 0.95$ and $\frac{\tau_c}{\tau} = 0.75$, in view that the maximum $P$ we observe is 0.82, and the maximum $C$ is 0.64. It should be noted that without these additional terms $C$ is always greater than $P$.